\documentclass[sigplan,10pt]{acmart}

\usepackage{tikz}
\usepackage{multirow}
\usepackage{amsmath}
\usepackage{subcaption}
\usepackage[linesnumbered,ruled,vlined]{algorithm2e}
\usepackage{xspace}
\usepackage{booktabs}
\usepackage{xcolor}

\copyrightyear{2026}
\acmYear{2026}
\setcopyright{cc}
\setcctype{by}
\acmConference[SOSP '26]
  {ACM SIGOPS 32nd Symposium on Operating Systems Principles}
  {September 29--October 02, 2026}
  {Prague, Czech Republic}
\acmBooktitle{ACM SIGOPS 32nd Symposium on Operating Systems Principles
  (SOSP '26), September 29--October 02, 2026, Prague, Czech Republic}
\acmDOI{10.1145/3830418.3843856}
\acmISBN{979-8-4007-2585-2/2026/09}

\begin{document}

\newcommand{\sys}{MeshRT\xspace}
\newcommand{\func}[1]{\textup{\textsc{#1}}}
\newcommand{\mypar}[1]{\par\smallskip\noindent\textbf{#1.}\xspace}

\newcommand*\myc[1]{\scalebox{0.78}{\begin{tikzpicture}[baseline=-3pt]
  \node[draw,circle,inner sep=0.5pt, fill=black] {\textcolor{white}{\textsf{\textbf{#1}}}};
\end{tikzpicture}}}

\newenvironment{tightlist}{
\begin{list}{$\bullet$}{
    \setlength{\topsep}{.1em}
    \setlength{\partopsep}{0in}
    \setlength{\parskip}{0in}
    \setlength{\itemsep}{0in}
    \setlength{\parsep}{0in}
    \setlength{\leftmargin}{1em}
    \setlength{\rightmargin}{0in}
    \setlength{\itemindent}{0in}
}}
{\end{list}}

\newenvironment{tightenum}{
\begin{list}{\myc{\arabic{enumi}}}{
    \usecounter{enumi}
    \setlength{\topsep}{.1em}
    \setlength{\partopsep}{0in}
    \setlength{\parskip}{0in}
    \setlength{\itemsep}{0in}
    \setlength{\parsep}{0in}
    \setlength{\leftmargin}{1.5em}
    \setlength{\rightmargin}{0in}
    \setlength{\itemindent}{0in}
}}
{\end{list}}

\setlength{\textfloatsep}{10pt plus 2pt minus 2pt}
\setlength{\floatsep}{8pt plus 2pt minus 2pt}

\title{MeshRT: Compile-Time Governed Wafer-Scale Runtime for Low-Latency High-Throughput Inference}

\makeatletter
\def\@ACM@checkaffil{}
\makeatother

\author{
{\rm Congjie He$^1$ \hspace{0.05cm} Le Xu$^1$ \hspace{0.05cm} Zhan Lu$^1$ \hspace{0.05cm} Yeqi Huang$^1$ \hspace{0.05cm} Haocheng Xiao$^1$ \hspace{0.05cm} Cheng Deng$^1$ \hspace{0.05cm} Lingxiao Ma$^2$ \hspace{0.05cm} Ziming Miao$^2$ \hspace{0.05cm} Fan Yang$^2$ \hspace{0.05cm} Luo Mai$^1$} \\
{\it $^1$University of Edinburgh \quad $^2$Microsoft Research}
}

\renewcommand{\shortauthors}{He et al.}

\begin{abstract}

Wafer-scale accelerators promise ultra-low-latency AI inference, but current system stacks still carry an unsustainably high cost premium. The reason is that many current and emerging inference techniques, such as batching and MoE, introduce runtime dynamism that existing wafer-scale systems cannot support efficiently. As a result, they must choose between two unsatisfactory options: accept suboptimal latency, or preserve low latency by restricting dynamic techniques, thereby limiting the models they can support and reducing aggregate throughput. This tradeoff significantly increases the cost of low-latency serving.

We present \sys{}, the first wafer-scale system to achieve both ultra-low latency and high throughput for LLM inference. \sys{} achieves this via a novel system architecture, which we call \emph{compile-time governance of runtime dynamism}. This architecture features new system abstractions and mechanisms that enable \sys{} to compile dynamic events arising from LLM inference directly into per-core schedules, enabling asynchronous and decentralized execution with low runtime overhead. Implemented on a commodity wafer-scale accelerator, \sys{} preserves ultra-low latency while improving throughput by up to 56$\times$ over the state-of-the-art system for wafer-scale accelerators. Against measured single-instance SGLang results on H200, \sys{} achieves 4.5--7.0$\times$ lower latency. Under ideal linear scaling of those GPU measurements to an approximately power-matched H200 cluster, \sys{} achieves 1.7--13.0$\times$ higher decode system throughput in the low-latency regime.

\end{abstract}



\settopmatter{printfolios=false}
\maketitle
\pagestyle{empty}

\section{Introduction}\label{sec:intro}

Low-latency LLM inference is becoming a core requirement where users and providers are willing to pay a premium for faster responses (\S\ref{sec:premium-market}). This drives a growing interest in inference infrastructure that can deliver substantially lower latency than conventional GPU clusters even with small batches.

Wafer-scale accelerators~\cite{wsdl2019} are especially promising for this setting. With massive on-chip SRAM and dense on-chip interconnect, they can reduce the off-chip overhead that dominates inference latency on conventional systems. Yet low-latency serving remains expensive at scale. In practice, providers often operate conservatively: they avoid aggressive batching, limit concurrency, and restrict dynamic inference features that could incur significant overhead. This preserves low latency, but it also leaves much of the hardware underutilized and makes low-latency serving far more expensive than throughput-oriented serving~\cite{he2025waferllm}.

This cost premium reflects a deeper systems limitation: modern inference workloads increasingly introduce \emph{runtime dynamism}, aspects of execution that the precompiled schedule cannot determine in advance and that must be resolved as inference proceeds. Common examples include changes in active request state during continuous batching or micro-batch pipelining, as well as token-dependent decisions such as Mixture-of-Experts (MoE) routing~\cite{fedus2022switch,dai2024deepseekmoe,jiang2024mixtral}, sparse attention~\cite{child2019generating,zhang2025spargeattention,chen2026msa,yuan2025native}, and speculative decoding~\cite{leviathan2023fast,miao2024specinfer}. On GPU systems, such runtime decisions are typically handled through host-managed runtime state~\cite{kwon2023vllm, zheng2024sglang}. On wafer-scale accelerators, however, sending them off chip or resolving them through na\"ive broad synchronization hurts both latency and throughput. Existing wafer-scale systems~\cite{he2025waferllm} struggle to provide support for runtime dynamism. This forces users into a no-win trade-off: either keep execution sufficiently static to preserve latency but limit system throughput, which leads to high serving cost, or support runtime dynamism at a high coordination overhead, hurting serving latency.

We present \sys{}, a system for wafer-scale inference that addresses this challenge through \emph{compile-time governance of runtime dynamism}: critical runtime decisions are governed by compile-time-generated on-chip schedules, rather than by falling back to host control or chip-wide synchronization. Here, "on-chip" refers to the compiled control state installed on the cores; \sys{}'s governance is unchanged whether model parameters and KV caches use SRAM alone or an SRAM--DRAM hierarchy. Our approach rests on a key observation: LLM inference uses a closed set of kernel types. Thus, although the specific values of runtime decisions are unknown until execution, the \emph{structure} of dynamism: where it arises, what type it is, and which cores are affected, is fully determinable at compile time. \sys{} exploits this by compiling dynamism handling directly into per-core schedules in a single synthesis pass, and then making these schedules executable under tight per-core resource budgets through safe on-chip reconfiguration. As a result, execution preserves the low-latency benefits of wafer-scale accelerators while supporting runtime dynamism and substantially improving throughput.

\sys{} makes two contributions. First, it introduces \emph{dynamism aware per-core schedule synthesis}: given an LLM computation graph and spatial mapping on a wafer-scale accelerator, the compiler determines, from each kernel's function alone, which cores will receive or produce runtime dynamism, and weaves the corresponding control message handling logic directly into the per-core schedule alongside ordinary data processing. This turns runtime dynamism from an external runtime responsibility into a compile-time scheduling object, providing on-chip runtime control for continuous batching, MoE routing, and micro-batch pipelining. Second, because the synthesized schedules may exceed the limited hardware resources, such as routing resources and queue buffers of individual cores, \sys{} introduces \emph{safe incremental materialization}: it first partitions each core's schedule into segments that fit within hardware budgets. Then it performs deadlock-free on-chip reconfiguration at segment boundaries, exploiting the all-or-nothing completion semantics of collectives to minimize synchronization cost.

We implement \sys{} on a commodity wafer-scale accelerator and evaluate it on widely used AI models, including both MoE models such as GPT-OSS 120B~\cite{gpt-oss-moe-120b}, GPT-OSS 20B~\cite{gpt-oss-moe-20b}, Qwen3.5-35B~\cite{qwen3-5-35b-a3b}, and dense models such as Qwen3.5-9B~\cite{qwen3-5-9b}. Compared to WaferLLM~\cite{he2025waferllm}, the state-of-the-art wafer-scale software stack, \sys{} adds 25--40\,$\mu$s (7.6--9.5\%) per decode token in matched configurations while improving throughput by up to 56$\times$. Against measured single-instance SGLang results on H200, \sys{} achieves 4.5--7.0$\times$ lower latency. Under ideal linear scaling of those GPU measurements to an approximately power-matched H200 cluster, \sys{} improves decode system throughput by 1.7--13.0$\times$ in the low-latency regime. We believe \sys{} marks an important step toward the coming era of low-latency AI, making such infrastructure substantially more affordable and broadly deployable.

\section{Background and Motivation}\label{sec:background}

\subsection{Low-Latency Inference as a Premium Market}\label{sec:premium-market}

Low-latency inference has become a first-class design objective. Interactive coding assistants~\cite{sergeyuk2025using, claude_code, codex, gemini_cli}, copilots~\cite{microsoft_copilot}, and agentic workloads~\cite{guo2024ds, gemini_deepresearch, chatgpt_deepresearch, openclaw} all depend on fast per-token generation: lower decode latency directly reduces round-trip time in multi-turn interactions and enables tighter feedback loops. Providers have begun offering dedicated low-latency serving tiers~\cite{aws_latency_optimized,anthropic_opus5_2026,openai_ultrafast_2026}, reflecting the fact that latency and throughput require fundamentally different system tradeoffs.

This demand is already reflected in production deployments. Anthropic offers a Fast mode for Claude Opus that runs at approximately 2.5$\times$ its default speed~\cite{anthropic_opus5_2026}. OpenAI serves Codex-Spark on Cerebras hardware at more than 1{,}000\,tokens/s for real-time coding workloads~\cite{openai_codex_spark_cerebras_2026} and previews an Ultrafast API tier serving GPT-5.6 Sol at up to 750\,tokens/s, up to 14$\times$ faster than standard processing~\cite{openai_ultrafast_2026}. Once low latency is productized as a premium tier, reducing its serving cost becomes an economic imperative, and the key lever is achieving high throughput without sacrificing the low latency that defines the tier.

\subsection{Wafer-Scale Chips for Low-Latency Inference}\label{sec:wse3}

\begin{figure}[t]
\centering
\includegraphics[width=0.95\linewidth]{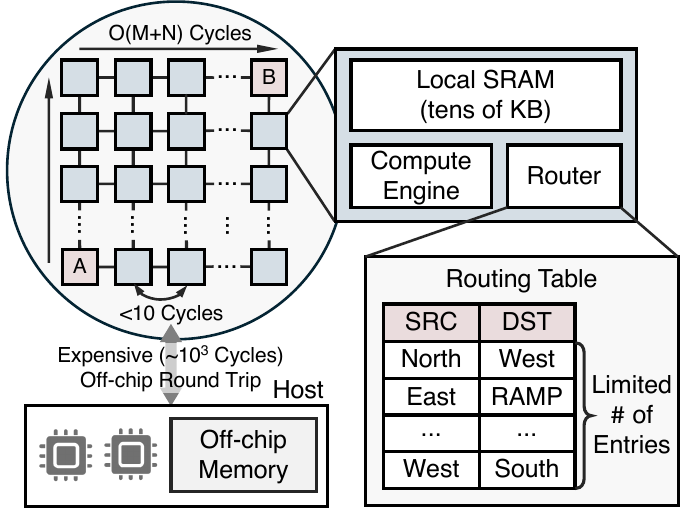}
\caption{Wafer-scale accelerators as a distinct runtime setting. Host-mediated control incurs expensive round trips. Cores form a neighbor-only mesh of $M\times N$ cores, so chip-wide synchronization takes $O(M{+}N)$ hops and grows with physical distance. Each core exposes only bounded local resources, including local SRAM, finite routing-table state, and simple local branching rather than a general scheduler. These properties limit design choices.}

\Description{A wafer-scale accelerator connected to a host. The accelerator contains an M-by-N neighbor-connected core mesh, making global communication proportional to the mesh dimensions. An enlarged core shows bounded local SRAM, finite routing state, and simple branching resources.}
\label{fig:wafer-scale-runtime}
\end{figure}

Wafer-scale accelerators (Figure~\ref{fig:wafer-scale-runtime}) are a natural fit for low-latency inference because they can reduce the two dominant sources of per-token latency on conventional hardware~\cite{he2025waferllm}. First, latency-critical data can reside in on-chip SRAM when capacity permits, reducing the off-chip HBM accesses that bottleneck decode on GPUs, where each token generation requires reading the full KV cache and model weights through a memory bus orders of magnitude slower than SRAM. Second, all inter-core communication traverses an on-chip network-on-chip (NoC) fabric rather than PCIe, NVLink, or InfiniBand, eliminating the off-chip interconnect latency that grows with tensor-parallel degree in multi-GPU deployments. The Cerebras WSE-3~\cite{cerebras2024wse3} exemplifies this architecture: it integrates approximately 900{,}000 processing elements (PEs) on a single wafer in a $762 \times 1172$ 2D array, with 44\,GB of aggregate on-chip SRAM and 21\,PB/s of aggregate memory bandwidth. Each PE (hereafter referred to as a core) has approximately 48\,KB of local SRAM, limited compute capability, and no shared memory; communication is point-to-point over the on-chip NoC connecting neighboring PEs~\cite{lie2023cerebras}.

Despite these latency advantages, the same architectural properties that reduce these bottlenecks also present the
following challenges for runtime design.

\mypar{Limited but programmable local control}
A processing core can execute simple conditional branches based on local data, but it cannot run an open-ended runtime scheduler. This is fundamentally different from a GPU SM, where general-purpose control logic can execute on chip. On wafer-scale hardware, runtime handling must therefore be compiled into bounded local logic rather than delegated to a rich scheduler.

\mypar{Expensive global coordination}
A wafer-scale accelerator has no shared memory, no global cache, and no cheap centralized coordination point visible to all cores. Cores communicate only through the on-chip network, and communication cost grows with physical distance on the wafer. As a result, any runtime design that repeatedly consults a global scheduler, a shared request queue, or any other chip-wide synchronization will be costly.

\mypar{Limited per-core resources to hold all execution states}
Each core has only a small amount of local SRAM and a bounded amount of routing and queue state. This rules out two tempting approaches: storing large off-chip-style runtime metainfo structures on chip, and materializing every possible execution configuration in advance. A practical runtime must instead operate within tight local budgets and support reconfiguration when the full schedule does not fit.

These constraints create a runtime setting that is fundamentally different from conventional shared-memory architectures like GPUs and static accelerator pipelines~\cite{yu2022orca,kwon2023vllm,ozkan2025performance}. Any runtime decision that GPU serving systems handle through a centralized host scheduler or shared memory must instead be resolved locally on each core, within tight resource budgets and without global coordination.

\subsection{Runtime Dynamism Erodes Latency}\label{sec:dynamism-taxonomy}

Modern LLM serving workloads introduce various forms of \emph{runtime dynamism}: aspects of the concrete execution path that cannot be fully determined by the precompiled schedule and must be resolved as inference proceeds.  Without explicit runtime support for on-chip dynamism, existing wafer-scale systems are forced into one of two costly fallbacks: falling back to static execution (sacrificing resource efficiency) or offloading dynamic decisions to the host (sacrificing latency).

\begin{table}[t]
  \centering
  \small
  \begin{tabular}{@{}lcl@{}}
  \toprule
  \textbf{Communication} & \textbf{Complexity} & \textbf{Cycle Cost} \\
  \midrule
  Broadcast              & $O(N)$              & $100-1000$  \\
  All-to-all             & $O(N^2)$            & $10^5 - 10^6$  \\
  Host synchronization   & $O(1)$              & $7000$  \\
  \bottomrule
  \end{tabular}
  \caption{Communication costs for runtime coordination on a wafer-scale accelerator.}
  \label{tab:comm-cost}
\end{table}

\mypar{Scenario 1: Static execution is wasteful due to dynamic request sets}
One important source of runtime dynamism is that the active request set evolves during execution. Continuous batching~\cite{yu2022orca} is a canonical example: to keep throughput high, the system must refill slots vacated by completed requests and determine which request state remains relevant at each decode step. Similar effects arise in micro-batch pipeline execution~\cite{oh2024exegpt,su2025seesaw}, dynamic batching~\cite{pang2025optimizing}, and speculative decoding with runtime acceptance decisions~\cite{leviathan2023fast,miao2024specinfer}.

On a wafer-scale chip, supporting this dynamism on chip presents a fundamental tension.  One option is to propagate request-completion events to all cores so that each can perform local management, for example, evicting stale KV cache entries or reassigning batch slots. But the chip lacks on-chip inter-core coordination: the only available mechanism is a blind broadcast that must reach all $\sim$900{,}000 cores, costing $O(N)$ hops and thousands of cycles per event (Table~\ref{tab:comm-cost}).  As batch size is crucial for high throughput serving, these operations accumulate quickly and inflate per-step latency.

\begin{figure}[t!]
    \centering
    \includegraphics[width=1.0\linewidth]{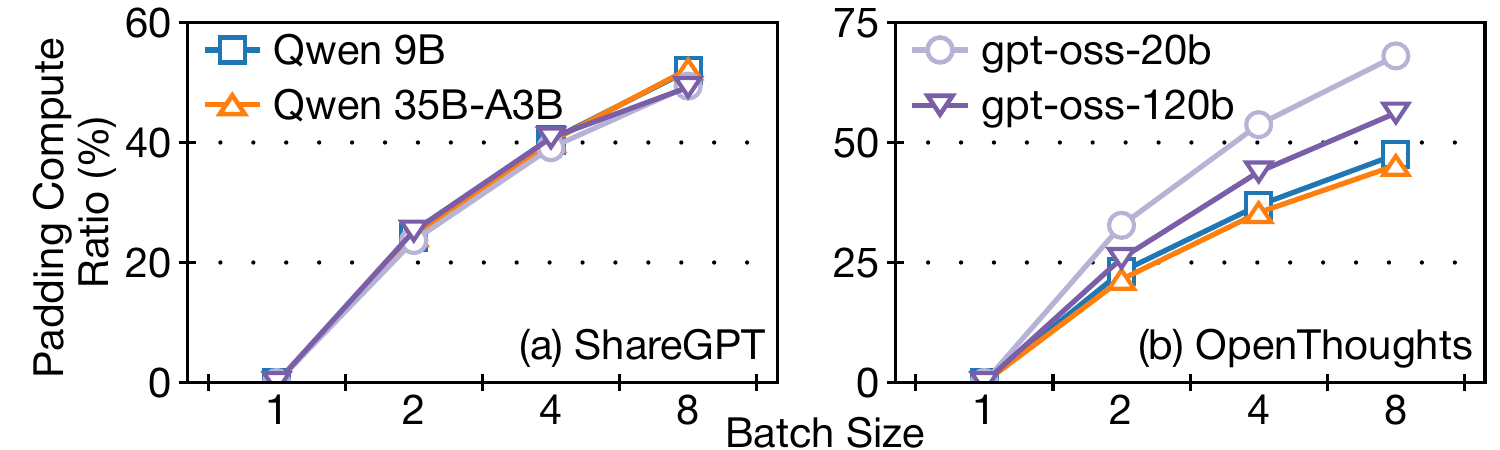}
    \caption{Padding compute ratios across different models increase with serving batch size.}
    \label{fig:sharegpt-padding-ratio}
    \Description{A plot of padding compute ratio versus serving batch size for the evaluated models on ShareGPT requests. Every model's padding ratio rises as batch size increases, showing that larger static batches waste more computation on padded tokens.}
\end{figure}

The alternative is to eliminate dynamism altogether by fixing the batch composition for the entire serving session, padding every request to the maximum sequence length so that no per-step reconfiguration is needed. This avoids broadcast overhead but trades it for wasted compute.

To quantify this waste, we measure the padding ratio, the fraction of decode compute spent on padded (zero-useful-work) tokens, under static batching across two representative workload distributions (ShareGPT~\cite{sharegpt} and
OpenThoughts~\cite{guha2025openthoughts}) and a mixture of dense and MoE models (Qwen3.5-Dense-9B, Qwen3.5-MoE-35B, GPT-OSS-MoE-20B, and GPT-OSS-MoE-120B). As depicted in Figure~\ref{fig:sharegpt-padding-ratio}, at batch size~8, padding consumes 45--52\% of compute on ShareGPT and 45--68\% on OpenThoughts. The problem is both consistent and severe: it holds across dense and MoE architectures alike, and worsens with higher output-length variance (OpenThoughts exhibits longer reasoning chains than ShareGPT).

\mypar{Scenario 2: Runtime decisions can stall the next step} A second source of runtime dynamism is that runtime results may determine what happens next, requiring coordinated reconfiguration of on-chip routing state. MoE serving with expert parallelism~\cite{rajbhandari2022deepspeed,fedus2022switch} is a concrete example: the gating network produces per-token expert assignments only after the current activation is processed, and these assignments vary from token to token. Similar effects arise in conditional-depth models~\cite{raposo2024mixture}, token-dependent routing~\cite{ainslie2023colt5}, and speculative decoding~\cite{leviathan2023fast}.

On a wafer-scale mesh, a layer's experts are partitioned across disjoint core groups so that the experts a step activates compute concurrently, each over a collective spanning far fewer cores. The group a token must reach is therefore known only after gating, and acting on this requires synchronized state changes: the attention-stage cores must communicate target-expert identifiers to the expert-stage cores so that both sides can configure routing entries for the correct data paths. Because these core groups may share relay links on the mesh, reconfiguration must be coordinated, installing a new routing entry while a neighbor still relies on the old one can corrupt in-flight data and cascade into deadlock. Without explicit on-chip support, the system faces two costly alternatives: all-to-all synchronization at $O(N^2)$ messages on a mesh with no multicast primitive (Table~\ref{tab:comm-cost}), or deferring each decision to the host at ${\sim}$7{,}000 cycles per round trip (Table~\ref{tab:comm-cost}), 1.9-3.5$\times$ the on-chip decode step itself. Since a single token generation involves hundreds of such interactions, host-mediated control accounts for 65-78\% of per-step latency in our measurements.

\subsection{Existing Approaches Fall Short}\label{sec:design-gap}

Existing systems largely make one of two design choices, and neither is satisfactory for low-latency wafer-scale inference.

\mypar{Approach 1: keep operator execution on chip, but keep scheduling logic off chip}
This is the dominant choice in GPU serving systems~\cite{kwon2023vllm,yu2022orca,zheng2024sglang}. Operators execute on the device, but request queues, dispatchers, and much of the runtime logic for handling dynamism remain on the host. This works well when the host can cheaply coordinate device execution and maintain large centralized runtime state.

On wafer-scale accelerators, however, this design choice conflicts with the hardware setting above. Repeated host intervention interrupts on-chip execution, incurs host-device round trips on latency-sensitive steps, and prevents the system from fully exploiting the benefit of keeping latency-sensitive control on chip. As a result, low latency becomes harder to maintain, and throughput suffers because runtime decisions cannot be absorbed cheaply where execution is happening.

\mypar{Approach 2: keep both operator execution and schedule on chip through precompiled execution}
This is the dominant choice in compiler-style wafer-scale and SRAM-first systems~\cite{abts2020thinkfast,lpu,he2025waferllm}: to preserve low latency, the system avoids off-chip runtime control by precompiling the execution schedule.

However, this design struggles with runtime dynamism. Continuous batching requires the active request set to evolve over time. MoE routing requires token-dependent decisions known only at runtime. Without an explicit mechanism for handling these decisions on chip, such systems either support these features poorly, restrict them heavily, or rely on expensive overprovisioning by precompiling many execution variants. On a wafer-scale accelerator, this extra state and routing cost is especially problematic because per-core resources are already tight. The result is a different but equally unsatisfactory outcome: low latency can be preserved, but throughput remains low and deployment cost remains high.

\section{Design Intuitions}\label{sec:design-intuitions}

\sys{} keeps the inference control path on chip and extends compiled execution so that important runtime events can also be handled on chip. We call this new system architecture approach \textbf{compile-time governance of runtime dynamism} on wafer-scale accelerators.

\begin{figure}[t]
    \centering
    \includegraphics[width=\linewidth]{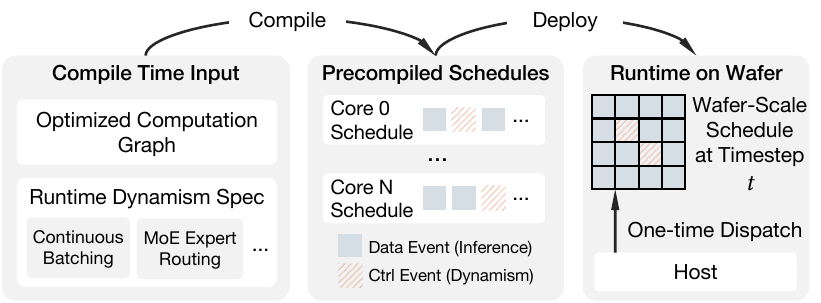}
    \caption{\sys{} design intuition: compile-time governance of runtime dynamism. Instead of deferring dynamic decisions to host control, \sys{} compiles them directly into per-core schedules as control events that execute alongside data events entirely on chip. Each core then runs its own schedule asynchronously, coordinating only through data arrival.}
    \label{fig:design-intuition}
    \Description{A comparison between conventional host-controlled execution and MeshRT. Conventional execution repeatedly sends dynamic decisions between the host and accelerator. MeshRT instead compiles both data and control events into per-core schedules; cores then execute asynchronously and coordinate through arriving data.}
\end{figure}

Figure~\ref{fig:design-intuition} illustrates this idea. \sys{} takes an optimized computation graph from an upstream compiler together with a specification of runtime dynamism (e.g., continuous batching, MoE routing) as input, and generates precompiled per-core schedules that interleave \emph{data events} for executing inference with \emph{dynamic control events} for handling runtime dynamism. These dynamic techniques are established inference optimizations, not contributions of this paper; what \sys{} contributes is making them applicable and efficient on a wafer-scale accelerator, where their usual implementations rely on a host runtime and shared memory that the hardware does not provide. Because both types of events are encoded at compile time, dynamic control remains on device: each core on the wafer-scale engine executes its schedule autonomously, host intervention is minimized, and the host only needs to install each per-core schedule once. The dynamism-aware precompiled schedule is then mapped and executed on the cores available on the wafer-scale accelerator. By doing so, \sys{} can preserve the low-latency benefits of wafer-scale accelerators while still supporting high-throughput features such as continuous batching and MoE routing.

Compile-time governance of runtime dynamism introduces two challenges to the wafer-scale system.

\mypar{Challenge 1: Dynamism-aware schedule synthesis (\S\ref{sec:design-synthesis})}
Because execution is compiled, every core needs a local schedule that specifies what to receive, compute, and send; collectively, all per-core schedules must faithfully realize the global computation graph. The input is a logical graph of kernel instances with global dependencies, not a set of local instruction streams. Bridging this gap already requires a non-trivial synthesis step; the additional requirement that runtime dynamism be absorbed \emph{within} the synthesized schedule, rather than deferred to a separate runtime layer, makes the problem harder. The compiler must identify, at synthesis time, which cores receive data whose interpretation depends on a runtime decision and which cores produce results that introduce new runtime decisions for downstream cores, and it must weave the corresponding control logic into the per-core schedule.

\mypar{Challenge 2: Safe incremental materialization (\S\ref{sec:design-materialization})}
To execute on chip without host intervention, schedules must fit within the limited memory, routing, and queue resources of each core. A core typically participates in many kernels across a forward pass, and the union of their resource footprints can exceed what the core can instantiate at once. On a GPU this is resolved by spilling to HBM or rescheduling at runtime; here, neither option is available. \sys{} must partition the schedule into segments and compile the segments as well as their transitions into deadlock-free on-chip execution, a common challenge in large-scale asynchronous distributed runtimes.

Both challenges rely on a shared on-chip control substrate (\S\ref{sec:design-synthesis}, \S\ref{sec:impl-datapath}). Each core maintains a small \emph{local control state}, such as which request slots are active or which routing result applies next, and uses bounded \emph{control messages} exchanged over the same on-chip network. For schedule synthesis, this substrate is what allows dynamism-handling logic to be woven into the compiled schedule: when the next data path is already valid, control metainfo rides alongside ordinary activations; when the path depends on a runtime result, a short control-message exchange resolves it within the affected region before data movement proceeds. For incremental materialization, the same substrate carries reconfiguration metainfo at segment boundaries, enabling cores to transition safely between materialized segments without host coordination.

Two properties make this substrate viable under the constraints of \S\ref{sec:wse3}. First, control messages are bounded: a few bytes of metainfo per batch slot, piggybacked on activations three orders of magnitude larger, so dynamism handling costs no separate network traversal. Second, participation is local: because the compiler already knows which cores a decision affects, only those cores exchange control messages, avoiding both the $O(N)$ broadcast and the host round trip priced in Table~\ref{tab:comm-cost}.

\section{System Design}\label{sec:design}

\begin{figure}[t]
    \centering
    \includegraphics[width=1.0\linewidth]{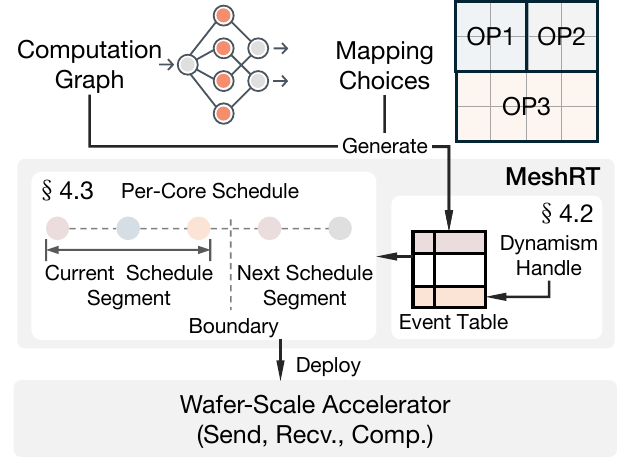}
    \caption{\sys{} system overview. Given a computation graph and mapping choices, \sys{} synthesizes dynamism-aware per-core event tables (\S\ref{sec:design-synthesis}), partitions them into segments that fit hardware resource limits (\S\ref{sec:design-materialization}), and deploys the result onto the wafer-scale accelerator.}
    \label{fig:system-overview}
    \Description{The MeshRT compilation and deployment pipeline. A computation graph and mapping choices enter schedule synthesis, which produces per-core event-action tables. Materialization partitions these tables into resource-fitting segments, and deployment installs the resulting schedules on the wafer-scale accelerator.}
\end{figure}

Figure~\ref{fig:system-overview} gives an overview of \sys{}. Given a computation graph, \sys{} synthesizes \emph{dynamism-aware per-core schedules} that interleave data-processing and dynamism-handling sub-schedules in a single compilation pass (\S\ref{sec:design-synthesis}). It then makes these schedules executable under tight per-core resource budgets through \emph{safe incremental materialization} (\S\ref{sec:design-materialization}).

\subsection{System Model and Design Assumptions}\label{sec:system-model}

\sys{} sits between an upstream compiler and the vendor-provided hardware primitives. Its input is a \emph{compiler-optimized computation graph} - a logical description of the inference workload.
Additionally, \sys{} requires \emph{mapping choices} that assign kernels, i.e., nodes in the computation graph, to spatial regions on the wafer.

A \emph{kernel instance} denotes a concrete occurrence of a kernel in the model's logical execution; each is associated with a kernel function that specifies its computation pattern and a mapping choice $R_k$ that assigns it to a spatial region on the wafer. The input to \sys{} is therefore a computation graph $G=(K,E)$ together with mapping choices $M=\{R_k \mid k \in K\}$, where $K$ is the set of kernel instances and $E$ is the set of logical dependencies between them. Edges in $E$ encode execution order: a kernel instance executes only after all of its predecessors have produced their outputs. Together, $G$ and $M$ define the mapped computation that \sys{} lowers to per-core execution.

The hardware exposes a per-core primitive interface $\Pi=\{\texttt{Compute}, \texttt{Receive}, \texttt{Send}\}$. \texttt{Compute} applies a local operation to core-resident data, including arithmetic computation, conditional branching, and local hardware reconfiguration (such as rebinding routing color IDs and buffer queue assignments on the WSE-3); \texttt{Receive} and \texttt{Send} move data between neighboring cores. \sys{} turns the mapped computation into per-core compiled schedules over this interface.

Note that mapping choices are important: prior work has explored mapping strategies for spatial accelerators~\cite{jiang2020cu,zhu2025theseus,kao2020gamma}. Such choices are outside our scope; we select them empirically for the evaluated models. Our focus is on the system design that follows once such choices are available.

\subsection{Dynamism-Aware Per-Core Schedule Synthesis}\label{sec:design-synthesis}

To synthesize per-core schedules that support runtime dynamism, our key observation is: although the specific values of inference runtime decisions are unknown until execution, the control \emph{structure} of a per-core schedule, including where dynamism arises, what type it is, and how each core should handle it, can be fully determined at compile time. LLM execution uses a set of \emph{known} kernel types. Once the graph and mapping are fixed, each kernel determines where dynamism may arise and which cores may consume or produce it. \sys{} captures this with two predicates, \func{ReceivesDynamism}$(k,c)$ and \func{ProducesDynamism}$(k,c)$. Whenever either holds, the compiler injects metainfo-handling sub-schedules around the kernel's core schedule. The injection is bounded: it applies only to the cores in $R_k$, the region the kernel is mapped to, so dynamism handling stays within the cores a decision actually reaches and never becomes chip-wide. In this way, \sys{} turns runtime dynamism from an external runtime responsibility into a compile-time scheduling object. Per-core resource limits are handled separately by incremental materialization (\S\ref{sec:design-materialization}).

This synthesis proceeds as shown in Algorithm~\ref{alg:schedule-synthesis}. (1)~\textbf{Initialization (line 1--2):} Allocate an empty schedule $S_c$ for each core that participates in at least one kernel instance. (2)~\textbf{Topological traversal (line 3--13):} Walk kernel instances in spatial-temporal dependency order so that each kernel is processed only after its predecessors are known. (3)~\textbf{Incoming dynamism resolution (line 6--8):} For each core $c$ in the kernel's assigned region $R_k$, the compiler checks whether $c$ receives runtime dynamism in kernel $k$ (\textsc{ReceivesDynamism}). If so, a metainfo resolution sub-schedule is inserted: the core first receives and parses incoming metainfo to resolve the dynamism before computation begins. (4)~\textbf{Kernel lowering (line 9--10):} The kernel's own core schedule is generated via \textsc{GetCoreSche}, determined by the kernel function $f_k$, the assigned region $R_k$, and the primitive interface $\Pi$. (5)~\textbf{Outgoing dynamism production (line 11--13):} The compiler checks whether $c$ produces runtime dynamism in kernel $k$ (\textsc{ProducesDynamism}). If so, a metainfo production sub-schedule is appended: the core generates metainfo from its computation result and sends it ahead of the data to downstream cores.

All sub-schedules are attached to $S_c$ via \textsc{AppendSche} together with the predecessor guard $p_k = \mathrm{Pred}(k)$. Iterating in topological order preserves the spatial-temporal dependencies of $G$, while iterating over $c \in R_k$ specializes the schedule to the spatial mapping $M$.

\begin{figure}[t]
    \centering
    \includegraphics[width=1.0\linewidth]{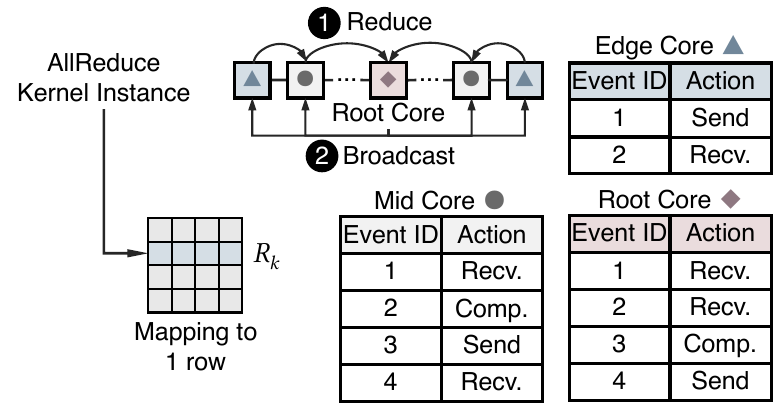}
    \caption{Per-core schedule synthesis for an AllReduce kernel mapped to one row of cores. The kernel function and mapping determine role-specific event-action tables: edge cores receive a partial sum, accumulate locally, forward it, and later receive the broadcast result; mid cores relay and accumulate; and the root reduces and broadcasts.}
    \label{fig:allreduce-schedule-synthesis}
    \Description{An AllReduce mapped across one row of cores, followed by event-action tables for each core role. Edge and middle cores receive, accumulate, and forward partial sums; the root core performs the final reduction and broadcasts the result back along the row.}
\end{figure}

Figure~\ref{fig:allreduce-schedule-synthesis} illustrates a synthesis example for \texttt{AllReduce}.
Consider an \texttt{AllReduce} kernel instance $k$ whose kernel function specifies a reduction over a set of cores, and whose mapping choice assigns it to a region $R_k$ of four cores arranged in a line. The compiler walks the cores in $R_k$ and calls \func{GetCoreSche}$(k, c)$ for each. The kernel function describes each core's role in both phases of the collective, so different cores receive different event-action tables. An edge core receives a partial sum, accumulates its local contribution, forwards the new partial sum toward the root, and later receives the broadcast result. A mid core relays and accumulates during reduction and relays the result in the reverse direction. The root receives the reduction contributions, computes the final result, and initiates the broadcast. The compiler also evaluates \func{ReceivesDynamism}$(k, c)$ and \func{ProducesDynamism}$(k, c)$ for each core; since \texttt{AllReduce}'s kernel function involves no runtime-dependent metainfo, both checks return false, and no metainfo sub-schedules are inserted. The result is a per-core schedule consisting purely of data-processing entries, the baseline case of Algorithm~\ref{alg:schedule-synthesis} where only step~(4) applies.

\begin{algorithm}[t]
\small
  \caption{Dynamism-aware Per-core Schedule}
  \label{alg:schedule-synthesis}
  \KwIn{$G = (K, E)$; $M = \{R_k \mid k \in K\}$; $\Pi$}
  \KwOut{$\{S_c\}$}
  \BlankLine
  \ForEach{$\func{Core}\; c \in \bigcup_{k \in K} R_k$}{
      $S_c \leftarrow [\ ]$\;
  }
  \BlankLine
  \ForEach{$k \in \func{TopologicalOrder}(G)$}{
      $p_k \leftarrow \mathrm{Pred}(k)$\;
      \ForEach{$c \in R_k$}{
          \If{$\func{ReceivesDynamism}(k, c)$}{
                $s_{\mathrm{recv}} \leftarrow \func{GetMetaResolver}(k, c)$\;
                \func{AppendSche}$(S_c,\; s_{\mathrm{recv}},\; p_k)$\;
            }
            $s_{k,c} \leftarrow \func{GetCoreSche}(k, c)$\;
            \func{AppendSche}$(S_c,\; s_{k,c},\; p_k)$\;
            \If{$\func{ProducesDynamism}(k, c)$}{
                $s_{\mathrm{send}} \leftarrow \func{GetMetaProducer}(k, c)$\;
                \func{AppendSche}$(S_c,\; s_{\mathrm{send}},\; p_k)$\;
            }
      }
  }
  \Return{$\{S_c\}$}\;
  \end{algorithm}

\begin{figure}[t]
    \centering
    \includegraphics[width=0.95\linewidth]{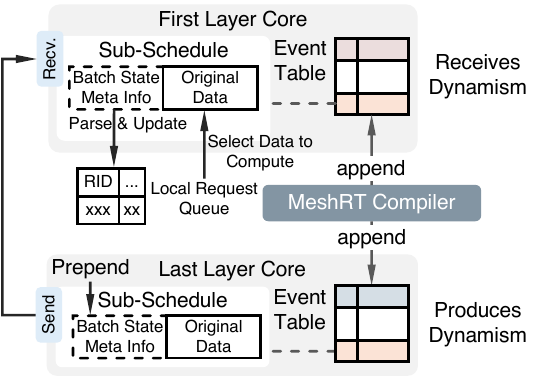}
    \caption{Synthesized per-core sub-schedules for continuous batching. The last-layer core prepends batch-state metainfo to outgoing data; the first-layer core parses it and updates its local request table to select active slots.}
    \label{fig:continuous-batching-resolution}
    \Description{A message-flow diagram between the last-layer and first-layer cores. The last-layer core attaches batch-state metadata to output data. The first-layer core parses the metadata, updates its local request table, and activates only the batch slots that should continue.}
\end{figure}

\begin{figure}[t]
    \centering
    \includegraphics[width=0.95\linewidth]{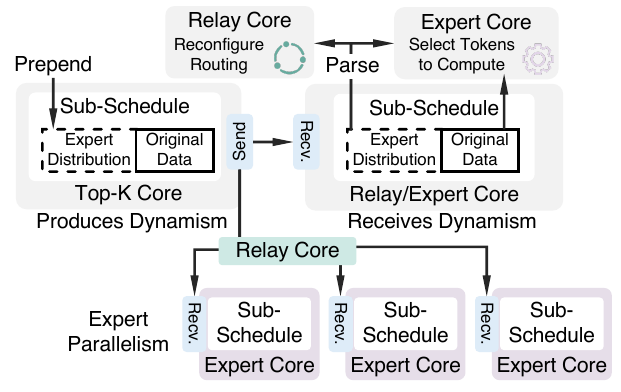}
    \caption{Synthesized per-core sub-schedules for MoE routing. The top-$k$ core prepends expert-distribution metainfo to token data; downstream relay cores parse it to reconfigure routing, and expert cores parse it to select which tokens to compute.}
    \label{fig:moe-routing-resolution}
    \Description{A message-flow diagram for mixture-of-experts routing. A top-k core attaches the selected expert distribution to token data. Relay cores read the metadata to configure token routes, and expert cores read it to select the tokens on which they operate.}
\end{figure}

\noindent \textbf{Use case: continuous batching} The active request set evolves during execution: requests complete at different decode steps and new requests must refill vacated slots, which batch slots are active cannot be fixed at compile time (Figure~\ref{fig:continuous-batching-resolution}). The sampling kernel at the final transformer layer satisfies \func{ProducesDynamism}: after computing output tokens, each core detects end-of-sequence events, generates batch-state metainfo encoding which slots have completed, and sends it ahead of data to the first layer of the next decode step. The first transformer layer satisfies \func{ReceivesDynamism}: each core parses this metainfo to learn which slots are active before computation begins. Request arrivals and completions are thus absorbed entirely on chip.
\mypar{Use case: MoE} Expert-parallel mappings, which \sys{} takes as input (\S\ref{sec:system-model}), place a layer's experts on disjoint core regions so that the experts a step activates compute concurrently. We call each spatial core region assigned to one expert or a fixed expert group an \emph{expert region}. The gating network selects a token's destination region only at runtime, so the attention-to-expert data paths cannot be fixed at compile time (Figure~\ref{fig:moe-routing-resolution}). The gating kernel from the attention stage satisfies \func{ProducesDynamism}: after computing top-$k$ expert selections, each core generates routing metainfo and prepends it to the token data sent toward the expert region. Relay cores along the path and the expert FFN kernel both satisfy \func{ReceivesDynamism}: relay cores parse routing metainfo to determine forwarding directions, and expert cores parse it to identify which tokens to process. Token-dependent routing is thus resolved hop-by-hop on chip: metainfo of compile-time-known size travels over compile-time-known candidate paths, establishing the MoE dispatch and combine routes at runtime.

\mypar{Generality beyond the two examples} Continuous batching and MoE routing are the two use cases we implement and evaluate, but the \func{ReceivesDynamism}/\func{ProducesDynamism} mechanism is general. A runtime decision fits the same pattern when its producing kernels emit bounded metainfo and downstream consumers use it to determine their next actions. For global hot-expert load balancing, local load-statistic kernels produce metainfo, a collective decision kernel receives these statistics and produces a load-balancing decision, and the affected routing and expert kernels receive that decision. Speculative decoding is one example: the verification kernel produces an acceptance decision (how many draft tokens to keep), and downstream kernels receive this decision to determine which token states to carry forward. Sparse or adaptive attention is another: a selection kernel produces a set of active KV entries, and the attention kernel receives this set to determine which entries to fetch and compute over. In each case, the compiler evaluates \func{ReceivesDynamism} and \func{ProducesDynamism} for every mapped kernel and core, and inserts the corresponding metainfo resolver and producer sub-schedules; no change to the synthesis algorithm is needed.

\subsection{Safe Incremental Schedule Materialization}\label{sec:design-materialization}

Once \sys{} has a precompiled schedule that includes both data-processing logic and event-driven logic for handling runtime dynamism (\S\ref{sec:design-synthesis}), the remaining problem is executing this schedule on cores with tight hardware budgets like routing resources. Our key idea is to partition the schedule into \emph{schedule segments} whose peak resource usage fits within hardware limits, and to materialize these segments incrementally at runtime.

Incremental materialization must rely entirely on local, asynchronous on-chip message exchange; otherwise, cores would stall frequently for synchronized segment reconfiguration, increasing per-step latency and reducing throughput. At the same time, asynchronous reconfiguration must be \emph{safe}: switching segments without coordination can leave neighboring cores in inconsistent states, leading to deadlock.

We introduce \emph{safe incremental schedule materialization}, which prevents reconfiguration-induced deadlock. It reconfigures segments through local asynchronous event processing and message passing, without global synchronization. We describe its key steps below and omit its formal declaration due to the page limit:

\begin{tightenum}
    \item \textbf{Resource initialization.}
    For each core $c$, initialize resource budget $R \leftarrow \mathrm{Cap}(c)$ representing the core's full hardware capacity, including routing channels, queue bindings, buffers, and local control-state storage.
    \item \textbf{Continuous segment extension.} Walk the sub-schedules in $S_c$ in order. As long as the next sub-schedule fits within $R$, append it to the current segment and deduct its resource demand from $R$.
    \item \textbf{Overflow and boundary classification.} When the next sub-schedule exceeds $R$, the current segment must be closed. The compiler examines the most recent kernel in which core $c$ participated: if that kernel's function already guarantees that all participating cores reach a consistent safe state upon completion (as collectives such as \texttt{AllReduce} naturally do), the boundary is safe and no extra coordination is needed.
    \item \textbf{Closing the segment.}
    At a safe boundary, the compiler appends a reconfiguration sub-schedule to close the current segment. If the boundary is not inherently safe, the compiler first inserts a local synchronization step among the participating cores to establish a consistent state, and then appends the reconfiguration.

    \item \textbf{Resource reset and new segment.}
    After reconfiguration releases the old bindings, reset $R \leftarrow \mathrm{Cap}(c)$ and start a new segment beginning with the sub-schedule that triggered the overflow.
\end{tightenum}

\mypar{Why collectives provide natural safe boundaries} The key property is that a collective such as \texttt{AllReduce} has an \emph{all-or-nothing completion} semantics: a core can only finish its local part of the collective after it has both contributed its own data and received the final result, which in turn requires every other participating core to have contributed. The collective's completion events therefore provide a consistent cut over its participating region. When the compiler selects this point as a segment boundary, \sys{} reuses that cut without additional coordination.

\mypar{Why incremental schedule reconfiguration is safe}
We view a segment transition as a distributed snapshot: a consistent cut, rather than a simultaneous global barrier~\cite{chandy1985distributed}. \sys{} establishes this cut in two ways. When a segment ends at a collective, the collective's completion semantics already provide a consistent cut over its participating region. Otherwise, the compiler inserts a bounded marker exchange within the compile-time-known affected region, and each core switches when it reaches its local cut. Both cases preserve the same invariants: the cut covers every route affected by the reconfiguration, old-segment traffic remains before the cut, and new-segment traffic begins after it. Therefore, no message can observe mixed old and new bindings, preventing reconfiguration-induced deadlock without chip-wide synchronization.

\section{System Implementation}\label{sec:implementation}

\begin{figure}[t]
    \centering
    \includegraphics[width=1\linewidth]{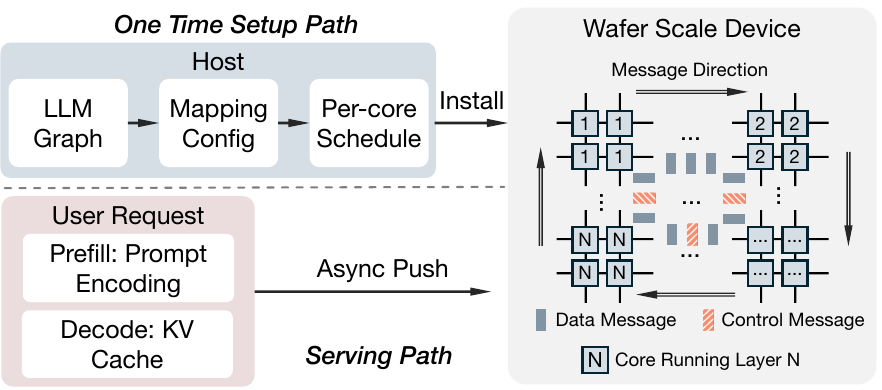}
    \caption{\sys{} system boundary. The host compiles and installs per-core schedules once (top left). During serving (bottom left), it asynchronously pushes prefill contexts and decode KV caches. All per-step execution, continuous batching, and dynamism resolution proceed on chip without host involvement (right).}
    \label{fig:system-implementation}
    \Description{A three-part architecture diagram. At upper left, the host compiler installs synthesized schedules on the wafer-scale accelerator. At lower left, the serving host asynchronously transfers prefill contexts and decode KV caches. On the right, a mesh of accelerator cores executes schedules, performs continuous batching, and resolves dynamic events entirely on chip.}
\end{figure}

We implement \sys{} on a Cerebras CS-3 system powered by the WSE-3 processor. Starting from a compiler-optimized computation graph and mapping choices, \sys{} synthesizes dynamism-aware per-core schedules (\S\ref{sec:design-synthesis}), partitions them via incremental materialization (\S\ref{sec:design-materialization}), and installs the resulting per-core executables onto the wafer once. Figure~\ref{fig:system-implementation} shows the resulting system boundary: the host performs one-time compilation and installation; during serving, it only pushes request data asynchronously. All per-step scheduling, batch management, and dynamism resolution proceed entirely on chip.

\subsection{Serving Data Path}\label{sec:impl-datapath}

Prefill processes prompts in chunks to reduce peak SRAM usage~\cite{agrawal2024sarathi}. Decode is where continuous batching operates. Each core maintains a \emph{local waiting request queue}, into which the host asynchronously pushes decode requests as (\texttt{request\_id}, KV~cache) or prefill requests as (\texttt{request\_id}, prompt token IDs), without interrupting on-chip execution, as well as a \emph{Request Descriptor Table~(RDT)} that records the current batch composition locally---including request identifiers, KV~cache pointers, sequence lengths, and other per-request metainfo. Micro-batches flow between pipeline stages carrying activation data and a compact piggybacked \emph{control-flow~(CF) payload}: only (\texttt{request\_id}, \texttt{status}) per batch slot, typically a few bytes per request versus ${\sim}$4\,KB of activation data ($<$0.2\% overhead).

\mypar{A decode step in action}
Consider slot~2 of a decoding micro-batch carrying request~17. While decode step~$t$ executes, the host may asynchronously enqueue request~42 without entering the per-token control path. At the final stage, on-chip sampling produces the next token; if request~17 emits an end-of-sequence token, the stage sets $\mathrm{CF}[2]=(17,\texttt{finished})$. At step~$t{+}1$, the first stage consumes this entry, releases request~17's KV slot, admits request~42 from its waiting queue, replaces $\mathrm{RDT}[2]=(17,\mathit{kv}_{17},\ell_{17})$ with $(42,\mathit{kv}_{42},\ell_{42})$, and emits $\mathrm{CF}[2]=(42,\texttt{active})$. Each following stage receives the activation and CF together, updates its local RDT, executes the ordinary layer events, and forwards both; sampled tokens in continuing slots feed back on chip to start the next autoregressive step. In the per-core event table, this adds one metainfo-resolution \texttt{Compute} before each affected layer and one metainfo-production \texttt{Compute} after sampling. Because CF is piggybacked on existing activation transfers, it adds no standalone \texttt{Send} or \texttt{Receive}. These events remain in the current materialized segment; segment transitions still occur only at the precompiled safe boundaries described in \S\ref{sec:design-materialization}.

\subsection{Multi-wafer Execution}
Our multi-wafer layouts partition the model only along the layer dimension: each wafer holds a contiguous stack of decoder layers and executes as one or more pipeline stages. We do not split an attention or expert layer across wafers, so KV state, expert routing, and schedule reconfiguration remain wafer-local. Consequently, the only bulk cross-wafer transfer is the hidden state at a pipeline-stage boundary; the CF payload rides with this hidden state and introduces no separate control round trip. Cerebras reports 1.2\,Tbps of inter-system bandwidth for CS-3~\cite{lie2024wafer}. In our measurements, these handoffs use approximately 100\,Gbps, less than 10\% of that bandwidth, and add less than 1\% to end-to-end decode latency.

\subsection{Safe Reconfiguration on GALS Hardware}\label{sec:impl-reconfig}

The WSE-3 is a globally asynchronous, locally synchronous (GALS) system: each core runs on its own clock and coordinates solely through data arrival. A core whose output queue has drained has reached \emph{local} quiescence, but upstream cores may still have data in transit through its router. Reconfiguring routing entries in this state corrupts in-flight data and causes deadlock. These races rarely surface in simulation, where the scale of parallelism is too small to expose meaningful asynchrony, and they manifest only at full hardware scale. Counterintuitively, low-latency inference maximizes this risk. Fine-grained parallelism means small per-core data volumes, which drain queues long before upstream cores finish, widening the race window. Large data transfers trigger backpressure that implicitly aligns cores; the minimal-data, maximal-parallelism regime \sys{} targets is the most exposed.

\sys{} applies two strategies from \S\ref{sec:design-materialization}. Most segment boundaries reuse the consistent cut already provided by a collective. For boundaries without such a natural cut, the expert-dispatch and combine sites of each MoE layer, \sys{} synthesizes bounded marker-based synchronization only within the affected region. The sequence of reconfiguration sites is fixed at compile time and repeats with each decode step; every transition remains local to one wafer, and pipeline-stage handoffs trigger no cross-wafer reconfiguration or synchronization. All reconfiguration costs are included in the reported device measurements.

\section{Evaluation}\label{sec:evaluation}

We evaluate \sys{} against WaferLLM~\cite{he2025waferllm}, the state-of-the-art in wafer-scale LLM inference systems, across four models spanning Dense and MoE architectures. Our evaluation answers three questions:

\begin{tightenum}
    \item Does \sys{}'s compile-time governance framework improve end-to-end inference throughput over static dataflow with branching? (\S\ref{sec:evaluation:e2e})
    \item How does \sys{} compare to GPU-based serving in latency, throughput, and energy efficiency? (\S\ref{sec:evaluation:gpu})
    \item What is the runtime cost of compile-time governance, and how do the serving policies it enables contribute to performance? (\S\ref{sec:evaluation:micro})
\end{tightenum}

\subsection{Experimental Setup}\label{sec:evaluation:setup}

\mypar{Testbed} Our wafer-scale testbed uses the Cerebras CS-3 system, powered by the WSE-3 processor: 900,000 compute cores, 44\,GB on-chip SRAM, 21\,PB/s memory bandwidth, and 214\,Pb/s core-to-core fabric bandwidth~\cite{cerebras2024wse3}. We use the vendor-reported 27\,kW system power figure as the budget per CS-3~\cite{cerebras2025cs3power}. For GPU comparisons, we use NVIDIA H200 SXM GPUs~\cite{nvidia2024h200} with SGLang 0.5.9~\cite{sglang-0-5-9} as the serving engine. For approximate power matching, we use 20 H200 GPUs per CS-3: an eight-GPU DGX H200 system has a maximum power rating of 10.2\,kW~\cite{nvidia2026dgxh200}, giving an aggregate node power budget of 25.5\,kW per 20 GPUs.\footnote{Compared with the SOSP version, we use approximate node power matching instead of chip-area matching for GPU throughput comparisons to reflect deployment power constraints. Power budgets directly constrain deployable serving capacity, whereas equal chip area does not account for differences in system-level power overheads.} These vendor-reported budgets cover compute nodes, excluding external infrastructure, and are not measured workload power. We measure the best-performing single data-parallelism instance on GPU (sweeping tensor-parallel degree), then linearly scale to 20 H200 GPUs per CS-3 in each model configuration.

\mypar{Models}
We evaluate four models that span the dense--MoE spectrum and exercise different dimensions of runtime dynamism:

\begin{tightlist}
      \item \textbf{Qwen3.5-Dense-9B}~\cite{qwen3-5-9b}: Dense transformer (9B parameters, FP16). Exercises continuous batching and pipeline parallelism but no expert routing. Runs on 2$\times$WSE-3.
      \item \textbf{Qwen3.5-MoE-35B}~\cite{qwen3-5-35b-a3b}: 35B-total, 3B-active MoE with 256 experts (top-8 + 1 shared), FP8 quantization. Requires a 6$\times$WSE-3 placement.
      \item \textbf{GPT-OSS-MoE-20B}~\cite{gpt-oss-moe-20b}: 20B-total, 3.6B-active MoE with 32 experts (top-4), MXFP4 quantization. Maps to 4$\times$WSE-3.
      \item \textbf{GPT-OSS-MoE-120B}~\cite{gpt-oss-moe-120b}: 120B-total, 5.1B-active MoE with 128 experts (top-4), MXFP4 quantization. Requires an 8$\times$WSE-3 placement.
\end{tightlist}

\mypar{Wafer-scale measurements}
For Qwen3.5-9B and GPT-OSS-20B, we measure the model-specific components needed by their complete 2- and 4-WSE-3 configurations, respectively, and compose full-model throughput from the measured device TSC cycles. These deployments span multiple wafers despite WSE-3's 44\,GB aggregate SRAM because the memory is distributed across PEs and each PE retains a complete compiled program image, even with staged schedule materialization, so that local sub-state machines can be activated at low overhead; the resulting replicated code footprint consumes substantial SRAM on every wafer. Keeping the program image resident does not eliminate the need for incremental materialization: only the active schedule segment instantiates its routing and queue state, since these per-core hardware resources remain scarce. For Qwen3.5-35B and GPT-OSS-120B, whose full placements span 6 and 8 WSE-3s, we measure their model-specific and transformer layer components on a two-wafer setup and project full-model results by composing the measured device TSC cycles according to each model's layer placement and pipeline-stage bottleneck. Unless stated otherwise, the reported 35B and 120B results use these projections. The approximately power-matched GPU cluster results are likewise projections: we measure the best-performing single data-parallel instance for each model and scale it ideally to the corresponding GPU count rather than measuring a physical multi-GPU cluster.

\mypar{Baseline} We compare against WaferLLM~\cite{he2025waferllm}, the published state of the art for LLM inference on the Cerebras WSE. WaferLLM uses PE-level if-else branching to select among pre-compiled execution paths, enabling multi-layer packing within a single PE block. However, it lacks three capabilities that \sys{} provides: (1)~per-request metainfo propagation (no continuous batching, batches are fixed with padding to uniform batch size), (2)~distributed routing resolution (no cross-block expert routing, all MoE experts are spatially tiled in a single PE block, forfeiting expert parallelism), and (3)~decentralized PE reconfiguration.

WaferLLM represents the best strategy achievable with PE-level branching alone, without a systematic governance framework built atop it. Both systems share the same static execution foundation (2D-tiled GEMM/GEMV, AllReduce via neighbor-based reduction trees); all differences stem from how each system handles runtime dynamism.

\mypar{Metrics}
We report: (1)~single-user throughput (tokens/s), the generation speed experienced by one user; (2)~system throughput (tokens/s), aggregate tokens produced across all concurrent requests; and (3)~estimated energy efficiency (tokens/joule), system throughput divided by the aggregate node power budget.

\mypar{Batching configurations}
For \sys{} on WSE-3, \texttt{micro-bs} denotes the number of requests
processed together within a pipeline stage. Models use different numbers
of wafers and spatial mappings, resulting in different pipeline stage
counts, which may also change with \texttt{micro-bs}. We therefore sweep
a common set of micro-batch sizes within each inference phase. The total
serving batch size is the in-flight request concurrency $c$, reported in
Figure~\ref{fig:end-to-end-pareto}. Under the evaluated pipeline
configurations, $c=\texttt{micro-bs}\times\text{stages}$.

\subsection{End-to-End Performance}\label{sec:evaluation:e2e}

\begin{figure*}[ht]
    \centering
    \includegraphics[width=0.9\linewidth]{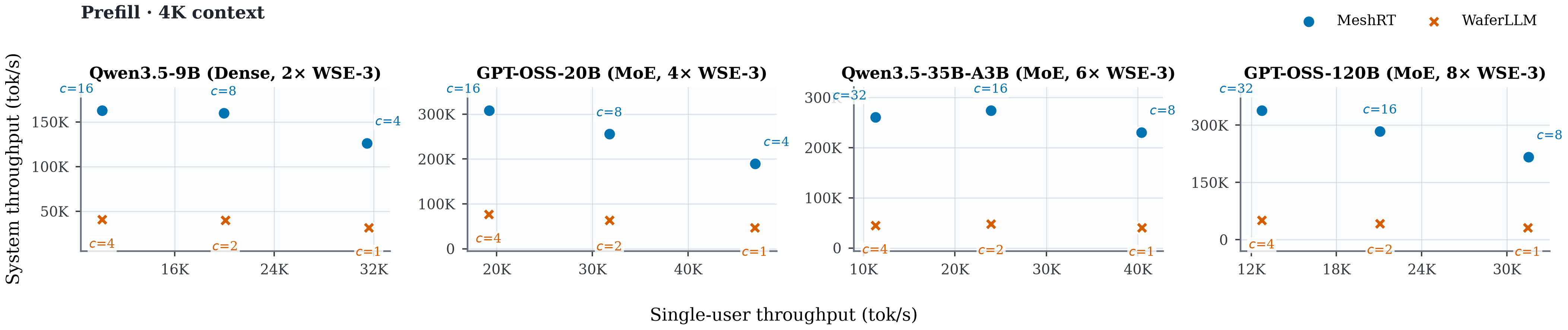}\\[3pt]
    \includegraphics[width=0.9\linewidth]{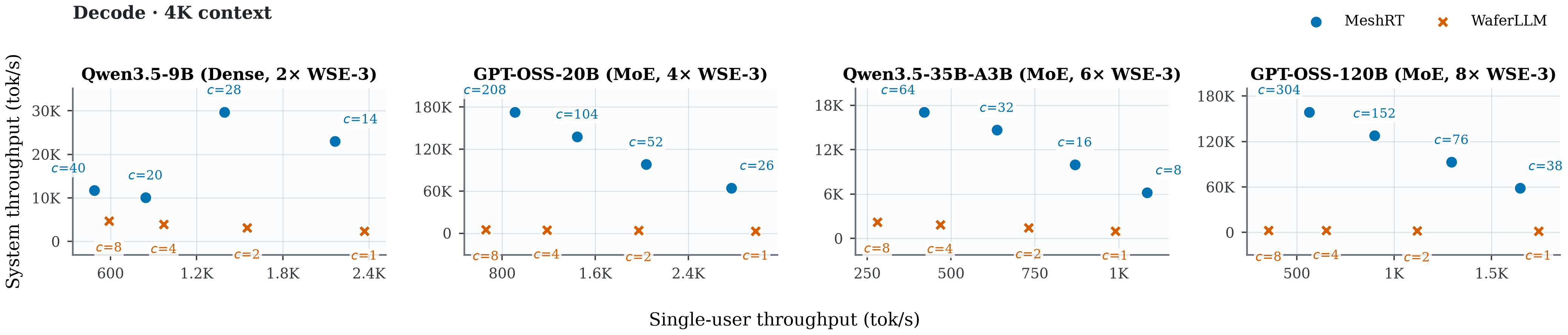}
    \caption{System throughput vs.\ single-user throughput on WSE-3 at 4K
    context for prefill (top) and decode (bottom). Comparing each system's
    peak over the evaluated micro-batches, \sys{} improves system throughput
    by 4.0--6.6$\times$ during prefill and by up to 55.8$\times$
    during decode. Decode points do not apply workload-level padding correction.}
    \label{fig:end-to-end-pareto}
    \Description{Eight scatter plots show system throughput against single-user throughput for four models during prefill and decode. Blue MeshRT circles form substantially higher throughput trade-off curves than orange WaferLLM crosses in every panel. Labels beside points give each evaluated concurrency, with lower concurrency generally moving right toward higher single-user throughput and lower aggregate throughput.}
\end{figure*}

This experiment compares the single-user and system throughput Pareto frontiers
of \sys{} and WaferLLM across all four models.

\mypar{Prefill experiments}
At 4K context, the two systems have nearly identical single-user throughput at each micro batch. Comparing the highest-system-throughput point of each system over the evaluated micro-batches, \sys{} improves system throughput by 4.0$\times$ on Qwen3.5-9B and GPT-OSS-20B, 5.7$\times$ on Qwen3.5-35B, and 6.6$\times$ on GPT-OSS-120B. The current MoE prefill variants use the same single-block FFN mapping as their static baselines. Table~\ref{tab:ablation} separately decomposes a controlled path from the static \texttt{bs}=1 configuration to pipelined \sys{} at \texttt{micro-bs}=1 and then to \texttt{micro-bs}=4; its factors therefore use different endpoints from these peak-to-peak ratios.

\mypar{Decode experiments} At the highest-system-throughput point for each model in Figure~\ref{fig:end-to-end-pareto}, \sys{} improves over WaferLLM by 6.3$\times$ on Qwen3.5-9B, 32.7$\times$ on GPT-OSS-20B, 7.6$\times$ on Qwen3.5-35B, and 55.8$\times$ on GPT-OSS-120B. These gains come from overlapping micro-batches across the compiled pipeline and, for MoE models, executing experts in parallel. The plotted values do not include workload-level padding correction; continuous batching's additional contribution is quantified in Table~\ref{tab:ablation}. The compact Qwen3.5-35B projection reported in Tables~\ref{tab:gpu-combined} and~\ref{tab:governance-overhead} is not part of this common-layout Pareto sweep.

\begin{figure}[t]
    \centering
    \includegraphics[width=\columnwidth]{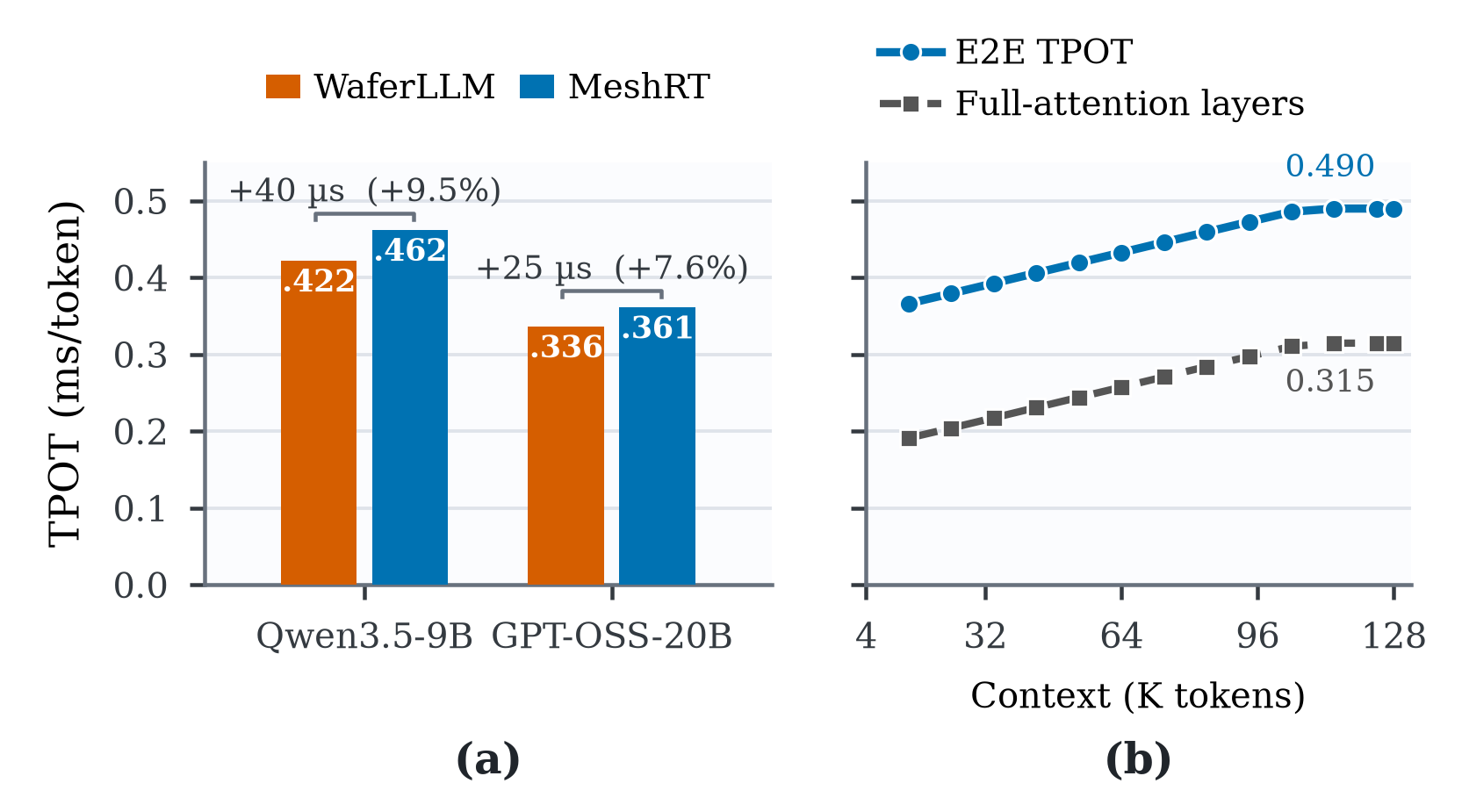}
    \caption{Decode TPOT (lower is better).
    (a) At 4K context and micro-batch size~1, under matched configurations, \sys{} adds only
    40\,$\mu$s (9.5\%) on Qwen3.5-9B and 25\,$\mu$s (7.6\%) on
    GPT-OSS-20B relative to WaferLLM. (b) At batch size~1, GPT-OSS-20B long-context decode,
    showing E2E TPOT and the time spent in full-attention layers as the
    context grows to 128K.}
    \label{fig:decode-long-context}
    \Description{Two decode-latency plots. The left grouped bar chart compares WaferLLM and MeshRT at 4K context: TPOT increases from 0.422 to 0.462 milliseconds for Qwen3.5-9B and from 0.336 to 0.361 milliseconds for GPT-OSS-20B. The right line chart shows GPT-OSS-20B latency from 16K to 128K context; end-to-end TPOT rises and then levels near 0.490 milliseconds, while full-attention time levels near 0.315 milliseconds.}
\end{figure}

\mypar{Governance overhead}
Figure~\ref{fig:decode-long-context}(a) compares per-user throughput for the two models with matched \sys{} and WaferLLM configurations. At 4K context and micro-batch size~1, \sys{} delivers 2{,}163 and 2{,}770\,tok/s per user on Qwen3.5-9B and GPT-OSS-20B, respectively, versus 2{,}368 and 2{,}980\,tok/s per user for WaferLLM. Equivalently, on-chip governance adds only 40\,$\mu$s (9.5\%) and 25\,$\mu$s (7.6\%) per token. In particular, the larger GPT-OSS-20B does not incur a larger penalty, showing that this overhead is determined by the fixed metainfo handling rather than model size. Because CF metainfo is piggybacked on activations, governance adds no standalone send or receive; continuous batching therefore increases throughput without adding a separate communication step to the per-token path.

Figure~\ref{fig:decode-long-context}(b) further shows that per-user throughput decreases gradually as the context grows toward 128K because full-attention work increases. Beyond approximately 103K context, new KV entries are transferred to the FFN PEs and processed in parallel with the fixed 431-slot attention prefix on the ATTN PEs. Since the FFN-side work has not yet overtaken the ATTN-side critical path, per-user throughput remains nearly flat thereafter.

\mypar{Spatial batching tradeoff}
Qwen3.5-9B exposes a wafer-scale-specific non-monotonicity. At \texttt{micro-bs}=2, each layer uses a $256{\times}256$-PE region, so the two WSE-3s fit 12 layer regions; including the head and tail yields 14 pipeline stages. At \texttt{micro-bs}=4 and~8, each layer instead requires a $512{\times}512$-PE region, leaving room for only three layer regions and thus five stages including the head and tail. The corresponding in-flight concurrencies, $c=\texttt{micro-bs}\times\text{stages}$, are $2\times14=28$, $4\times5=20$, and $8\times5=40$. System throughput is not this product: it is \texttt{micro-bs} divided by the bottleneck-stage time. That time rises from 74K cycles at \texttt{micro-bs}=2 to 436K and 752K cycles at \texttt{micro-bs}=4 and~8, yielding 29.7K, 10.1K, and 11.7K\,tok/s, respectively. Thus, the slower, coarser stages outweigh the benefit of a larger micro batch.

\subsection{Comparison with GPU Serving}\label{sec:evaluation:gpu}

We compare \sys{} on WSE-3 against SGLang on H200 GPUs under latency-oriented and throughput-oriented policies for both prefill and decode (Table~\ref{tab:gpu-combined}). Prefill uses 4K context length. GPU system throughput assumes ideal linear scale-out to 20 H200 GPUs per CS-3 and is therefore an optimistic upper bound.

\newcommand{\rg}[1]{\textcolor{green!60!black}{#1}}
\newcommand{\rb}[1]{\textcolor{red!70!black}{#1}}
\begin{table}[t]
  \centering
  \begin{subtable}{\columnwidth}
  \centering
  \footnotesize
  \setlength{\tabcolsep}{2pt}
  \begin{tabular*}{\columnwidth}{@{\extracolsep{\fill}}ll rrrrrr@{}}
  \toprule
  & &
  \multicolumn{3}{c}{\textbf{Prefill}} &
  \multicolumn{3}{c}{\textbf{Decode}} \\
  \cmidrule(lr){3-5} \cmidrule(lr){6-8}
  \textbf{Model} & &
  \textbf{User} & \textbf{Sys} & \textbf{tok/J} &
  \textbf{User} & \textbf{Sys} & \textbf{tok/J} \\
  \midrule
  \multirow{3}{*}{\shortstack[l]{Qwen3.5\\-9B}}
  & GPU   & 72{,}757 & 728K & 14.26 & 355     & 1{,}774  & 0.035 \\
  & \sys{}& 31{,}448 & 126K     & 2.33  & 2{,}163 & 22{,}990 & 0.426 \\
  & Ratio & \rb{0.43$\times$} & \rb{0.17$\times$} & \rb{0.16$\times$}
          & \rg{6.1$\times$}  & \rg{13.0$\times$}  & \rg{12.2$\times$} \\
  \midrule
  \multirow{3}{*}{\shortstack[l]{GPT-OSS\\-20B}}
  & GPU   & 80{,}264 & 3{,}211K & 31.48 & 393     & 7{,}859 & 0.077 \\
  & \sys{}& 47{,}018 & 190K     & 1.76  & 2{,}770 & 64{,}622 & 0.598 \\
  & Ratio & \rb{0.59$\times$} & \rb{0.06$\times$} & \rb{0.06$\times$}
          & \rg{7.0$\times$}  & \rg{8.2$\times$}  & \rg{7.8$\times$} \\
  \midrule
  \multirow{3}{*}{\shortstack[l]{Qwen3.5\\-35B-A3B}}
  & GPU   & 53{,}745 & 1{,}613K & 10.54 & 243     & 3{,}640 & 0.024 \\
  & \sys{}& 40{,}389 & 231K     & 1.43  & 1{,}085 & 6{,}207  & 0.038 \\
  & Ratio & \rb{0.75$\times$} & \rb{0.14$\times$} & \rb{0.14$\times$}
          & \rg{4.5$\times$}  & \rg{1.7$\times$} & \rg{1.6$\times$} \\
  \midrule
  \multirow{3}{*}{\shortstack[l]{GPT-OSS\\-120B}}
  & GPU   & 56{,}473 & 2{,}259K & 11.07 & 276     & 5{,}523 & 0.027 \\
  & \sys{}& 31{,}518 & 217K     & 1.00  & 1{,}647 & 58{,}305 & 0.270 \\
  & Ratio & \rb{0.56$\times$} & \rb{0.10$\times$} & \rb{0.09$\times$}
          & \rg{6.0$\times$}  & \rg{10.6$\times$}  & \rg{10.0$\times$} \\
  \bottomrule
  \end{tabular*}
  \subcaption{Latency-oriented.
  GPU uses each model's latency-optimal \texttt{tp}.}
  \label{tab:gpu-latency}
  \end{subtable}

  \begin{subtable}{\columnwidth}
  \centering
  \footnotesize
  \setlength{\tabcolsep}{2pt}
  \begin{tabular*}{\columnwidth}{@{\extracolsep{\fill}}ll rrrrrr@{}}
  \toprule
  & &
  \multicolumn{3}{c}{\textbf{Prefill}} &
  \multicolumn{3}{c}{\textbf{Decode}} \\
  \cmidrule(lr){3-5} \cmidrule(lr){6-8}
  \textbf{Model} & &
  \textbf{User} & \textbf{Sys} & \textbf{tok/J} &
  \textbf{User} & \textbf{Sys} & \textbf{tok/J} \\
  \midrule
  \multirow{3}{*}{\shortstack[l]{Qwen3.5\\-9B}}
  & GPU   & 9{,}167  & 1{,}467K & 28.8  & 165     & 13{,}193  & 0.26  \\
  & \sys{}& 10{,}161 & 163K     & 3.02  & 1{,}393 & 29{,}686  & 0.55  \\
  & Ratio & \rg{1.1$\times$}  & \rb{0.11$\times$} & \rb{0.10$\times$}
          & \rg{8.4$\times$}  & \rg{2.3$\times$} & \rg{2.1$\times$} \\
  \midrule
  \multirow{3}{*}{\shortstack[l]{GPT-OSS\\-20B}}
  & GPU   & 18{,}499 & 5{,}920K & 58.0 & 193     & 123{,}454 & 1.21  \\
  & \sys{}& 19{,}188 & 308K      & 2.85  & 910     & 172{,}614 & 1.60  \\
  & Ratio & \rg{1.0$\times$}  & \rb{0.05$\times$} & \rb{0.05$\times$}
          & \rg{4.7$\times$}  & \rg{1.4$\times$} & \rg{1.3$\times$} \\
  \midrule
  \multirow{3}{*}{\shortstack[l]{Qwen3.5\\-35B-A3B}}
  & GPU   & 22{,}016 & 5{,}284K & 34.5  & 114     & 109{,}888 & 0.72  \\
  & \sys{}& 11{,}289 & 261K      & 1.61  & 419     & 63{,}274  & 0.39  \\
  & Ratio & \rb{0.51$\times$} & \rb{0.05$\times$} & \rb{0.05$\times$}
          & \rg{3.7$\times$}  & \rb{0.58$\times$} & \rb{0.54$\times$} \\
  \midrule
  \multirow{3}{*}{\shortstack[l]{GPT-OSS\\-120B}}
  & GPU   & 11{,}523 & 7{,}375K & 36.2  & 107     & 137{,}536 & 0.67  \\
  & \sys{}& 12{,}753 & 338K      & 1.56  & 564     & 158{,}701 & 0.73  \\
  & Ratio & \rg{1.1$\times$}  & \rb{0.05$\times$} & \rb{0.04$\times$}
          & \rg{5.3$\times$}  & \rg{1.2$\times$} & \rg{1.1$\times$} \\
  \bottomrule
  \end{tabular*}
  \subcaption{Throughput-oriented.
  GPU system throughput projected by linearly scaling \texttt{tp}=1 (upper bound).}
  \label{tab:gpu-throughput}
  \end{subtable}
  \caption{\sys{} vs.\ GPU (H200) across prefill and decode.
  \textbf{User} and \textbf{Sys} denote per-user and aggregate throughput.
  Each block reports GPU, \sys{}, and the \sys{}/GPU ratio
  (\rg{green} ${\geq}1{\times}$, \rb{red} ${<}1{\times}$).
  GPU system throughput is projected to 20 H200 GPUs per CS-3.
  The 35B and 120B \sys{} results are projected as described in
  \S\ref{sec:evaluation:setup}. Energy estimates use vendor-reported system power budgets of 27\,kW/CS-3 and 10.2\,kW/DGX H200 (eight GPUs), not measured workload power.
  All throughputs in tok/s; energy in tok/J.}
  \label{tab:gpu-combined}
  \end{table}

\mypar{Decode: \sys{} excels in per-user latency and throughput} For latency-oriented decode (\texttt{micro-bs}=1), \sys{} improves per-user throughput by 4.5--7.0$\times$ and system throughput by 1.7--13.0$\times$; estimated energy efficiency improves by 1.6--12.2$\times$ over the GPU projection. GPU scale-out raises aggregate throughput but does not improve the per-user speed of each data-parallel instance. \sys{}'s per-user advantage comes from the WSE-3's 21\,PB/s SRAM bandwidth and from on-chip dynamism resolution via CF~payloads and RDTs (\S\ref{sec:impl-datapath}), which avoids per-step host--device coordination.

For throughput-oriented decode (Table~\ref{tab:gpu-throughput}), \sys{} retains a 3.7--8.4$\times$ per-user advantage, while delivering 1.2--2.3$\times$ higher aggregate throughput on three models and 0.58$\times$ on Qwen3.5-35B under ideal GPU scaling. \sys{} remains 1.1--2.1$\times$ more energy-efficient on Qwen3.5-9B, GPT-OSS-20B, and GPT-OSS-120B, while reaching 0.54$\times$ on Qwen3.5-35B. These operating points expose the tradeoff: \sys{} consistently improves request-level decode speed, while its aggregate-throughput advantage depends on the model.

\mypar{Prefill: structural mismatch, mitigated in practice} Prefill is compute-bound (GEMM), and here the WSE-3's architecture is at a structural disadvantage: its 900{,}000 simple cores trade compute density for aggregate SRAM bandwidth, whereas the H200's tensor cores are purpose-built for high-FLOPS matrix multiplication. For latency-oriented prefill, GPU per-user throughput is 1.3--2.3$\times$ higher; under throughput-oriented operation, projected GPU system throughput exceeds \sys{} by 9--22$\times$ under ideal linear scaling.

Pipeline parallelism partially compensates at the request level. At \texttt{micro-bs}=4, overlapping requests across pipeline stages allows \sys{} to match or slightly exceed GPU per-user throughput on Qwen3.5-9B, GPT-OSS-20B, and GPT-OSS-120B (1.0--1.1$\times$). Qwen3.5-35B reaches 0.51$\times$ because its per-user throughput falls beyond \texttt{micro-bs}=2. The aggregate-throughput gap remains large for all four models.

\subsection{Microbenchmarks}\label{sec:evaluation:micro}

\subsubsection{Cost of Compile-Time Governance}

\begin{table}[t]
    \centering
    \small
    \setlength{\tabcolsep}{2pt}
    \resizebox{\columnwidth}{!}{ \begin{tabular}{@{}l cc c cc c@{}}
    \toprule
    & \multicolumn{3}{c}{\textbf{Single-User (tok/s)}} & \multicolumn{3}{c}{\textbf{System (tok/s)}} \\
    \cmidrule(lr){2-4} \cmidrule(lr){5-7}
    \textbf{Model} & WaferLLM & MeshRT & Ratio & WaferLLM & MeshRT & Ratio \\
    \midrule
    Qwen3.5-9B       & 2{,}368 & 2{,}163 & \rb{0.91$\times$} & 4{,}717  & 29{,}686  & \rg{6.3$\times$} \\
    GPT-OSS-20B      & 2{,}980 & 2{,}770 & \rb{0.93$\times$} & 5{,}272  & 172{,}614 & \rg{32.7$\times$} \\
    \midrule
    Qwen3.5-35B-A3B  & 990     & 1{,}085 & \rg{1.10$\times$} & 2{,}238  & 63{,}274  & \rg{28.3$\times$} \\
    GPT-OSS-120B     & 1{,}742 & 1{,}647 & \rb{0.95$\times$} & 2{,}843  & 158{,}701 & \rg{55.8$\times$} \\
    \bottomrule
    \end{tabular} }
\caption{Matched-configuration per-step throughput and peak system throughput relative to static WaferLLM. Single-user throughput uses \texttt{micro-bs}=1 for \sys{} and \texttt{bs}=1 for WaferLLM; runtime governance latency overhead for the matched-configuration 9B and 20B comparisons is reported separately in Figure~\ref{fig:decode-long-context}. System throughput is each system's peak over the evaluated micro-batches and includes the effect of enabled policies. The 35B and 120B results follow the projection methodology in \S\ref{sec:evaluation:setup}. No workload-level padding correction is applied.}
\label{tab:governance-overhead}
\end{table}

Table~\ref{tab:governance-overhead} separates matched-configuration per-step throughput from the peak system throughput enabled by compile-time governance. At \texttt{micro-bs}=1, Qwen3.5-9B and GPT-OSS-20B retain 0.91--0.93$\times$ the baseline's single-user throughput under identical mappings; the corresponding governance overhead is 25--40\,$\mu$s, or 7.6--9.5\%, of TPOT in Figure~\ref{fig:decode-long-context}. We use TPOT for governance-overhead claims because it directly measures the latency added to each decode step. The Qwen3.5-35B-A3B and GPT-OSS-120B rows include configuration changes and are not governance-only A/Bs.

The system columns answer the complementary policy-level question by reporting the best aggregate throughput for each system. Across these operating points, the expert-parallel and dynamically pipelined execution enabled by compile-time governance improves peak system throughput by 6.3--55.8$\times$.

\subsubsection{Ablation Study}\label{sec:evaluation:ablation}

\begin{table}[t]
    \centering
    \footnotesize
    \begin{tabular*}{\columnwidth}{@{\extracolsep{\fill}}l rrrr@{}}
    \toprule
    \multicolumn{5}{@{}l}{\emph{Decode (\texttt{micro-bs}=4)}} \\
    \textbf{Model} & \textbf{Expert} & \textbf{+ Pipe.} & \textbf{+ Cont.} & \textbf{Total} \\
    \midrule
    GPT-OSS-20B     & 1.2$\times$ & 20.0$\times$ & 1.6$\times$ & \textbf{40.8$\times$} \\
    Qwen3.5-35B     & 1.2$\times$ & 5.5$\times$  & 1.7$\times$ & \textbf{10.7$\times$} \\
    GPT-OSS-120B    & 1.4$\times$ & 30.1$\times$ & 1.7$\times$ & \textbf{71.3$\times$} \\
    \midrule
    \multicolumn{5}{@{}l}{\emph{Prefill (seq\,=\,4096)}} \\
    \textbf{Model} & \textbf{Pipe. (\texttt{mbs}=1)} & \textbf{+ Req. (\texttt{mbs}=4)}
                   & \multicolumn{2}{r@{}}{\textbf{Total ($\texttt{mbs}:1\!\to\!4$)}} \\
    \midrule
    GPT-OSS-20B     & 4.0$\times$ & 1.6$\times$ & \multicolumn{2}{r@{}}{\textbf{6.6$\times$}} \\
    Qwen3.5-35B     & 5.7$\times$ & 1.1$\times$ & \multicolumn{2}{r@{}}{\textbf{6.4$\times$}} \\
    GPT-OSS-120B    & 6.9$\times$ & 1.6$\times$ & \multicolumn{2}{r@{}}{\textbf{10.8$\times$}} \\
    \bottomrule
    \end{tabular*}
    \caption{Policy-level decomposition of serving performance.
    In the table headers, \texttt{mbs} denotes micro-batch size. Each non-total
    column reports the multiplier over the preceding configuration after adding
    the named policy; the total is their product using unrounded values.
    Policy factors are derived from device-measured component cycles. Decode
    additionally uses the measured ShareGPT padding ratios for continuous
    batching. The prefill block reports the three MoE models: its first factor
    compares pipelined \sys{} at \texttt{micro-bs}=1 with the static \texttt{bs}=1
    baseline, and the second factor moves \sys{} from \texttt{micro-bs}=1 to
    \texttt{micro-bs}=4. The total therefore differs
    from Figure~\ref{fig:end-to-end-pareto}, which compares each system's peak
    over all evaluated micro-batches.}
    \label{tab:ablation}
\end{table}

Where Table~\ref{tab:governance-overhead} quantifies the cost of the
governance substrate, Table~\ref{tab:ablation} decomposes performance gains at
the policy level. Dynamism-aware schedule synthesis and safe incremental
materialization provide the substrate on which these policies execute. Decode
adds expert parallelism, pipeline scheduling, and continuous batching. The
current prefill implementation uses the same single-block FFN mapping as its
static baseline, so its columns report only pipeline scheduling and request
batching.

For decode, pipeline parallelism is the dominant factor
(5.5--30.1$\times$), followed by continuous batching (1.6--1.7$\times$)
and expert parallelism (1.2--1.4$\times$). Without continuous batching,
completed requests are padded until all requests in the same micro-batch
finish. Continuous batching's factor at \texttt{micro-bs}=4 is a
conservative lower bound and grows with micro-batch size.
For prefill on the three MoE models, pipeline scheduling at \texttt{micro-bs}=1 contributes
4.0--6.9$\times$. Increasing \sys{} from \texttt{micro-bs}=1 to \texttt{micro-bs}=4
adds 1.1--1.6$\times$, for cumulative gains of 6.4--10.8$\times$ over the
static \texttt{bs}=1 baseline. These controlled cumulative gains are distinct
from Figure~\ref{fig:end-to-end-pareto}'s 4.0--6.6$\times$ 4K peak-to-peak
range.

\section{Related Work}\label{sec:related}

\mypar{Deterministic dataflow accelerators}
The Groq TSP~\cite{abts2020thinkfast, lpu}, Graphcore IPU~\cite{jia2019dissecting}, and SambaNova RDU~\cite{prabhakar2024sambanova} all pre-compile execution schedules with no runtime scheduling overhead. The shared limitation is that any runtime uncertainty -- such as variable batch size, MoE routing, request arrival / departure, forces execution off the fast path (host round-trip on IPU, unsupported on Groq). \sys{} occupies a different point: it pre-compiles per-core schedules that \emph{govern} these dynamic events on-chip, preserving zero-runtime-scheduling while supporting continuous batching and expert routing. T10~\cite{liu2024t10} compiles individual DL operators onto IPU-like inter-core processors via the rTensor abstraction; \sys{} extends to full-model per-core schedules that handle cross-operator runtime dynamism.

MoEntwine~\cite{tang2026moentwine} optimizes large-scale expert parallelism on wafer-scale chips, including expert placement, all-to-all congestion, and replica migration for load balancing. In contrast, \sys{} takes spatial mappings as input and provides a general on-chip control-plane mechanism for runtime dynamism, including MoE routing; the two approaches are therefore potentially composable.
Tesseract~\cite{ahn2015scalable} is a programmable processing-in-memory accelerator that sends non-blocking remote function calls through message queues to the cores that own graph data, placing control requests alongside distributed data processing. It targets latency hiding and atomic updates in PIM graph processing rather than safe reconfiguration of a dynamically evolving compiled schedule on a GALS wafer. In contrast, \sys{} compiles runtime-dynamism handling into per-core schedules and safely reconfigures incrementally materialized segments to fit each core's bounded routing and buffering resources.

\mypar{GPU persistent execution}
Mirage~\cite{wu2025mirage} compiles tensor programs into mega-kernels via multi-level superoptimization, and CUDA Graphs~\cite{nvidia2018cudagraphs} pre-record kernel sequences for replay---both reduce launch overhead on GPUs. Mirage is structurally the most comparable to \sys{}'s always-on core execution. The key difference: GPU SMs support JIT dispatch and dynamic rebalancing via general-purpose control flow; WSE-3 PEs cannot, so \sys{} must resolve all scheduling at compile time.

\mypar{LLM serving systems}
Orca~\cite{yu2022orca} introduces continuous batching; vLLM~\cite{kwon2023vllm} adds paged KV cache management; SGLang~\cite{zheng2024sglang} enables KV cache reuse via RadixAttention; TensorRT-LLM~\cite{tensorrt-llm} provides production GPU kernels with in-flight batching. All assume a centralized host scheduler and shared GPU memory. \sys{} achieves continuous batching on mesh hardware where neither exists, using embedded control metainfo in compile-time schedules. DistServe~\cite{zhong2024distserve} and Splitwise~\cite{patel2024splitwise} disaggregate prefill and decode onto separate GPU pools; Sarathi-Serve~\cite{agrawal2024sarathi} co-locates them with chunked prefill; Mooncake~\cite{qin2025mooncake} disaggregates KV cache storage. LoongServe~\cite{wu2024loongserve} dynamically adjusts sequence parallelism for long contexts.

\mypar{DL compilers and parallelism frameworks}
Ansor~\cite{zheng2020ansor}, Roller~\cite{zhu2022roller}, and Triton~\cite{tillet2019triton} generate high-performance tensor programs for single-device shared-memory GPUs. Megatron-LM~\cite{narayanan2021megatron}, Alpa~\cite{zheng2022alpa}, FlexFlow~\cite{jia2019flexflow}, and DeepSpeed-Inference~\cite{aminabadi2022deepspeed} partition computation across discrete GPUs with GBs of HBM each, coordinated by a centralized scheduler over NVLink / InfiniBand. \sys{} targets a fundamentally different substrate: ${\sim}$900K PEs with 48\,KB each on a 2D mesh NoC, a three-order-of-magnitude difference in device count and per-device memory that demands a different approach to compilation, parallelism, and communication.

\mypar{Wafer-scale and mesh architectures}
The Cerebras WSE series~\cite{lie2023cerebras, lauterbach2021cerebras} provides the hardware substrate. WaferLLM~\cite{he2025waferllm} is the first published WSE inference system, using PE-level if-else branching among pre-compiled paths. \sys{} replaces branching with compile-time schedules that express continuous batching, expert-parallel MoE routing, and pipeline parallelism, capabilities beyond WaferLLM's branching model.

\section{Discussion and Limitations}\label{sec:discussion}

\mypar{PE fault tolerance}
\sys{} relies on the vendor-provided manufacturing-defect masking and remapping that uses redundant cores and routing to present a functional mesh~\cite{lauterbach2021cerebras}. Beyond that hardware layer, \sys{} does not detect or recover from PE failures that occur at runtime.

\mypar{Multi-tenancy}
\sys{} assumes exclusive access to the PE mesh for a single model. Serving multiple models concurrently on the same wafer: partitioning PE regions across models and potentially migrating them introduces resource management challenges not addressed by the current design.

\mypar{Dynamic model loading}
The current design assumes model weights are loaded onto PEs before inference begins and remain fixed. Hot-swapping model weights (\emph{e.g}.\null{} for A/B testing or model updates) would require additional mechanisms for coordinated weight replacement.

\mypar{Closed operator space assumption}
\sys{}'s completeness argument depends on the assumption that transformer inference decomposes into a finite, known set of mesh kernel primitives. Novel operator types (\emph{e.g}.\null{} state-space models, linear attention variants) may require new kernel implementations and potentially new dynamism-handling mechanisms.

\mypar{Globally coupled decisions and dependency stalls}
Dynamism aware synthesis represents both local and globally coupled runtime decisions in the generated per-core schedules. For each mapped kernel and core, \func{ReceivesDynamism} and \func{ProducesDynamism} determine whether the compiler inserts a metainfo resolver before the kernel's data-processing sub-schedule or a metainfo producer after it. A global policy such as hot-expert load balancing is encoded as a chain of local load-statistic producers, a collective decision kernel, and downstream consumers; the same synthesis pass generates the role-specific schedules for all participants. This does not eliminate the collective's intrinsic aggregation cost. Moreover, a consumer's schedule must receive and install the required metainfo before dependent data movement begins; a delayed decision can therefore stall that fragment, and the compiled schedule cannot dynamically reorder unrelated work unless the overlap was encoded at compile time.

\section{Conclusion}\label{sec:conclusion}

Wafer-scale accelerators offer compelling latency advantages for LLM inference, but existing software stacks sacrifice utilization by avoiding runtime dynamism. \sys{} resolves this through compile-time governance: for each compiled model, the compiler uses its finite, statically known set of mapped kernels to determine where dynamism arises and weave handling logic directly into per-core schedules, while safe incremental materialization ensures these schedules execute within tight hardware resource budgets. The result is a fully decentralized on-chip control path that preserves wafer-scale latency benefits while supporting continuous batching, MoE routing, and micro-batch pipelining. On Cerebras CS-3 systems, \sys{} improves throughput by up to 56$\times$ over WaferLLM while adding 25--40\,$\mu$s (7.6--9.5\%) per decode token in matched configurations. Against measured single-instance SGLang results on H200, \sys{} delivers 4.5--7.0$\times$ lower decode latency; with ideal linear scaling to an approximately power-matched H200 cluster, it achieves up to 13.0$\times$ higher decode system throughput.

\bibliographystyle{ACM-Reference-Format}
\bibliography{main}

\end{document}